\documentclass[sigconf]{aamas}
\setcopyright{none}
\renewcommand\footnotetextcopyrightpermission[1]{}
\usepackage{balance} 
\usepackage{subcaption}
\usepackage{dsfont}
\usepackage{geometry, graphicx}
\usepackage{amsmath}
\usepackage{algorithm}
\usepackage[noend]{algpseudocode}

\acmSubmissionID{???}

\title{Searching with Opponent-Awareness}

\author{Timy Phan}
\affiliation{
  \department{Department of Computer Science and Mathematics}
  \institution{Munich University of Applied Sciences}
  \city{Munich}
  \country{Germany}
}
\email{phantimy1999@gmail.com}

\begin{abstract}
We propose \emph{Searching with Opponent-Awareness (SOA)}, an approach to leverage opponent-aware planning without explicit or a priori opponent models for improving performance and social welfare in multi-agent systems. To this end, we develop an opponent-aware MCTS scheme using multi-armed bandits based on Learning with Opponent-Learning Awareness (LOLA) and compare its effectiveness with other bandits, including UCB1. Our evaluations include several different settings and show the benefits of SOA are especially evident with increasing number of agents.
\end{abstract}

\newcommand{\BibTeX}{\rm B\kern-.05em{\sc i\kern-.025em b}\kern-.08em\TeX}

\begin{document}


\pagestyle{fancy}
\fancyhead{}


\maketitle 


\section{Introduction}

In recent years, artificial intelligence methods have led to significant advances in several areas, especially reinforcement learning \citep{dqn, alphago, alphagozero, alphazero, muzero, alphastar}. Among these methods, planning is notable for exploiting the option of predicting possible directions into which a scenario can evolve. Planning has played a key role in achieving state-of-the-art performance in challenging domains like Atari, shogi, chess, go and hex \cite{alphago, alphagozero, alphazero, muzero, anthony}. Monte-Carlo planning enables scalable decision making via stochastic sampling by using a generative model as black box simulator instead of explicit environment models, where state and reward distributions need to be specified a priori \cite{mcplan}.

In a multi-agent system, individual agents have to make decisions in environments with higher complexity w.r.t. the number of dynamic elements (i.e. other agents). Agents may need to coordinate or to compete to satisfy their interests which results in a balance of conflict and cooperation. In a real-world use-case of multi-agent systems, like self-driving cars or workers in an industrial factory, agents are required to coordinate to avoid collisions and yet reach their individual goals. These interactions are studied in game theory and usually lead to a Nash equilibrium \cite{nashequilibrium, masfound, myersongametheory}.

Because the actions of each agent affect the payoffs of all other agents, it makes sense for agents to have an opponent model or to be opponent-aware in order to optimize their individual decisions. A naive agent only considers its own interests which can lead to traffic jams or accidents in the case of autonomous cars when conflicting interests escalate. An \emph{opponent model} is a representation of another agent which is given or obtained via \emph{opponent modelling}. \emph{Opponent-awareness} refers to an agent's ability to consider opponents in its own updates, either through direct access to opponent parameters or through approximation with opponent modelling.

Although modelling an opponent in planning is not a recent discovery \cite{distplanom, aamoa, instancebasedactionmodelsforfastactionplanning}, there is currently no approach which actually leverages opponent-awareness to improve social interaction between agents. The difficulty of opponent-aware planning in general-sum games lies in the lack of available and adequate opponent models in addition to the environment model. Special cases like cooperative or competitive games permit solutions which alleviate this problem by additional maximization or minimization operations \cite{mimimaxq, friendorfoeq, alphago, alphagozero}, but general-sum games become intractable w.r.t. the number of agents and possible outcomes.

In this paper, we propose \emph{Searching with Opponent-Awareness (SOA)} as an approach to opponent-aware planning without explicit and a priori opponent models in the context of Monte-Carlo planning. Our contributions are as follows:

\begin{itemize}
	\item A novel type of gradient-based multi-armed bandit which leverages opponent-awareness based on Learning with Opponent-Learning Awareness (LOLA) \cite{lola}.
	\item A novel MCTS variant using opponent-aware bandits to consider agent behavior during planning without explicit or a priori opponent models.
	\item An evaluation on several general-sum games w.r.t. to social behavior, payoffs, scalability and a comparison with alternative MCTS approaches. We analyze the behavior of planning agents in three iterated matrix games and two gridworld tasks with sequential social dilemmas and find that opponent-aware planners are more likely to cooperate. This leads to higher social welfare and we show that this benefit is especially evident with increasing number of agents when compared to naive planning agents.

\end{itemize}

\section{Related Work}

Opponent-awareness has been studied extensively in model-free reinforcement learning literature \cite{carmel, zhangom}, notably minimax-Q-learning \cite{mimimaxq}, Friend-or-Foe Q-learning \cite{friendorfoeq}, policy hill climbing \cite{winorlearnfast, zhangpolicyhill} and neural replicator dynamics \cite{neuralreplicatordynamics}. In particular, \emph{Learning with Opponent-Learning Awareness (LOLA)} exploits the update of a naive-learning opponent to achieve higher performance and social welfare \cite{lola}. Further work enhance LOLA learners through higher-order estimates of learning updates \cite{dice} and stabilization of fixed points \cite{sos}. All of these approaches are applied to model-free reinforcement learning, whereas in this paper we study model-based approaches and apply opponent-awareness to planning agents using search.

Model-based state-of-the-art approaches \cite{alphago, alphagozero, alphazero, muzero} also model opponents, though their application is restricted to zero-sum games. They are based on UCB1 bandits which select actions deterministically. This is insufficient in general-sum games where Nash equilibria generally consist of mixed strategies that select actions stochastically, as opposed to pure strategies which are deterministic. Furthermore, although these bandits are used to model opponents, the bandits themselves are not opponent-aware and their purpose is not to encourage cooperation or to optimize social welfare in general-sum games. In contrast, our proposed approach models opponents through \emph{opponent-aware} multi-armed bandits.

Though approaches like GraWoLF \cite{grawolf} and fictitious play \cite{ficplay} are applicable to general-sum games and take opponent behavior into account, note that these are \textit{learning} algorithms as opposed to our proposed \textit{planning} method.

Another way to implement opponent-awareness is through inter-agent communication which has also been studied in deep reinforcement learning \cite{learncomm, pitfallcomm} and planning \cite{wucomm, wuscience}. This is fundamentally different from using opponent modelling which is done decentralized by an individual agent and without assuming a separate channel for information exchange between agents \cite{lola, maddpg}.

\section{Notation}

We consider Markov games \cite{mimimaxq, sigaudbuffet} $\mathcal{M} = (n, \mathcal{S}, \mathcal{A}, \mathcal{P}, \mathcal{R})$ where $n$ is the number of agents, $i \in \{1, ..., n\}$, $\mathcal{S}$ is the set of states, $\mathcal{A} = \mathcal{A}_1 \times ... \times \mathcal{A}_n$ is the set of joint actions, $\mathcal{P}(s_{t+1}|s_t, a_t)$ is the transition probability function and $\mathcal{R}(s_t, a_t) = \mathcal{R}_1(s_t, a_t) \times ... \times \mathcal{R}_n(s_t, a_t)$ is the joint reward function. In this paper, we assume that $s_t, s_{t+1} \in \mathcal{S}$ and $a_t \in \mathcal{A}$ always holds and $t \in \mathbb{N}_0$ is a given time step. The joint action at time step $t$ is given by $a_t$ and the individual action of agent $i$ is $a_{i,t}$.

The behavior of agent $i$ is given by its policy $\pi_i : \mathcal{S} \times \mathcal{A}_i \rightarrow [0,1]$ with $\sum_{a_i \in \mathcal{A}_i} \pi(a_i|s) = 1$ and the joint policy of all agents is defined by $\pi = \langle \pi_1, ..., \pi_n \rangle$. The goal is to maximize the expectation of the discounted return $G_{i,t}$ at any given state.

\begin{equation}\label{eq:discountedreturn}
G_{i,t} = \sum_{k=0}^{h-1} \gamma^k \cdot \mathcal{R}_{i,t} (s_{t+k}, a_{t+k})
\end{equation}

where $h$ is the future horizon and $\gamma \in [0, 1]$ is the discount factor. The joint discounted return is $G_t$.

We measure social welfare with the collective undiscounted return $W$:

\begin{equation}\label{eq:return:undiscounted:collective}
W = \sum_{k=0}^{h-1} \sum_{i=1}^{n} \mathcal{R}_{i,t} (s_{t+k}, a_{t+k})
\end{equation}

A policy $\pi_i$ is evaluated with a state value function $V_i(s_t, \pi) = \mathbb{E}[G_{i,t} | s_t, \pi]$, which is defined by the expected discounted return at any $s_t$ \cite{bellman, boutilier} and the current joint policy $\pi$. The joint state value function $V = \langle V_1, ..., V_n \rangle$ evaluates the joint policy $\pi$. If a policy $\pi_i^{*}$ has a state value function $V_i^{*}$ where $V_i^{*} (s_t, \pi') \geq V_i' (s_t, \pi')$ for any $\pi_i'$ and $s_t \in \mathcal{S}$, given that $\pi_j'$ are constant for all $j \in \{1, ..., n\} \setminus i$, $\pi_i^{*}$ is a \emph{best response} to all $\pi_j'$ \cite{masfound, myersongametheory}. If $\pi_i$ is a best response for all $i \in \{1, ..., n\}$, $\pi$ is a \emph{Nash equilibrium} \cite{nashequilibrium, masfound, myersongametheory}. 

In game theory, the term \emph{strategy} is used to describe an agent's choice of actions at any given state \cite{masfound, myersongametheory}. There is a distinction between \emph{mixed strategies} and \emph{pure strategies} whereby the former type of strategy is stochastic and the latter is deterministic w.r.t. to the state \cite{masfound, myersongametheory}. In our case, an agent's mixed strategy is equivalent to its policy.

\section{Methods}
\subsection{Planning}
Planning searches for approximate best responses, given a generative model $\mathcal{\hat{M}}$ which approximates the actual environment $\mathcal{M}$, including $\mathcal{P}$ and $\mathcal{R}$. We assume $\mathcal{\hat{M}} = \mathcal{M}$. Global planning searches the whole state space for a global policy while local planning only considers the current state and possible future states within a horizon $h$ to find a local policy. We focus on local planning for online planning, i.e. actions are planned and executed at each time step given a fixed computation budget $c_b$.

\subsection{Multi-armed bandits}

\emph{Multi-armed bandits (MAB)} are agents used to solve decision-making problems with a single state $\mathcal{S}_\textit{MAB} = \{s_\textit{MAB}\}$. A bandit repeatedly selects one action or arm $a_\textit{MAB} \in \mathcal{A}_\textit{MAB}$ and is given a reward $\mathcal{R}_\textit{MAB} (s_\textit{MAB}, a_\textit{MAB}) = X_a$. $\mathcal{A}_\textit{MAB}$ is the set of available actions and $X_a$ is a stochastic variable of an unknown distribution. The goal is to maximize a bandit's expected reward $\mathbb{E}[X]$ by estimating the mean reward $\bar{X}_a$ for all $a_\textit{MAB} \in \mathcal{A}_\textit{MAB}$ and selecting $a_\textit{MAB}$ as $argmax (\bar{X}_a)$.

To learn the optimal MAB policy, a balance between selecting different arms to estimate $\bar{X}_a$ and selecting $argmax (\bar{X}_a)$ is needed. Finding such a balance is known as the \emph{exploration-exploitation dilemma}, where MAB algorithms trade-off avoidance of poor local optima and convergence speed.

\subsubsection{Upper Confidence Bound (UCB1)}

UCB1 bandits are commonly used in state-of-the-art approaches to MAB problems \cite{alphago, alphagozero, muzero, kocsis, bubeck}. The UCB1 criterion for selecting an action $a_{MAB}$ is given by \cite{auer}:

\begin{eqnarray}\label{eq:ucb}
UCB1_{a} = \bar{X}_a + C \cdot \sqrt{\frac{2 \log(N)}{N_{a}}}
\end{eqnarray}

where $\textit{C}$ is an exploration constant, $N$ is the iteration count and $N_{a}$ is the number of times $a_\textit{MAB}$ has been selected. UCB1 bandits always choose $a_\textit{MAB}$ through $argmax(UCB1_{a})$, though if there is more than one such $a_\textit{MAB}$, $a_\textit{MAB}$ is chosen from $argmax(UCB1_{a})$ randomly with uniform probability.

\subsubsection{Gradient-based (GRAB)}

GRAB bandits optimize $\mathbb{E}[X]$ through a variant of stochastic gradient ascent \cite{sutton}. Each $a_\textit{MAB}$ is assigned a numerical preference $H_N(a_\textit{MAB})$ ($N$ is the iteration count) and the probability $P_N(a_\textit{MAB})$ of selecting $a_\textit{MAB}$ is \cite{sutton}:

\begin{eqnarray}\label{eq:grabprob}
P_N(a_\textit{MAB})&=&\frac{e^{H_N(a_\textit{MAB})}}{\sum_{\tilde{a}_\textit{MAB} \in \mathcal{A}_\textit{MAB}}^{} e^{H_N(\tilde{a}_\textit{MAB})}}
\end{eqnarray}

$H_N$ is updated according to the following rule \cite{sutton}:

\begin{equation}\label{eq:grabupdate}
\begin{aligned}
H_{N+1} (\tilde{a}_\textit{MAB}) ={} & H_N(\tilde{a}_\textit{MAB}) \\
      & + \alpha (\bar{X}_a - X) (\mathds{1}_{\tilde{a}_\textit{MAB}=a_\textit{MAB}} - P_N(\tilde{a}_\textit{MAB}))
\end{aligned}
\end{equation}

$a_\textit{MAB}$ is the selected arm, $\alpha$ is the learning rate, $X$ is the current mean reward of this bandit, $\tilde{a}_\textit{MAB} \in \mathcal{A}_\textit{MAB}$ and $\mathds{1}_{\tilde{a}_\textit{MAB}=a_\textit{MAB}}$ is 1 iff $\tilde{a}_\textit{MAB}=a_\textit{MAB}$ and 0 otherwise.

\subsection{MCTS}

\begin{figure}
  \centering
  \includegraphics[width=\linewidth]{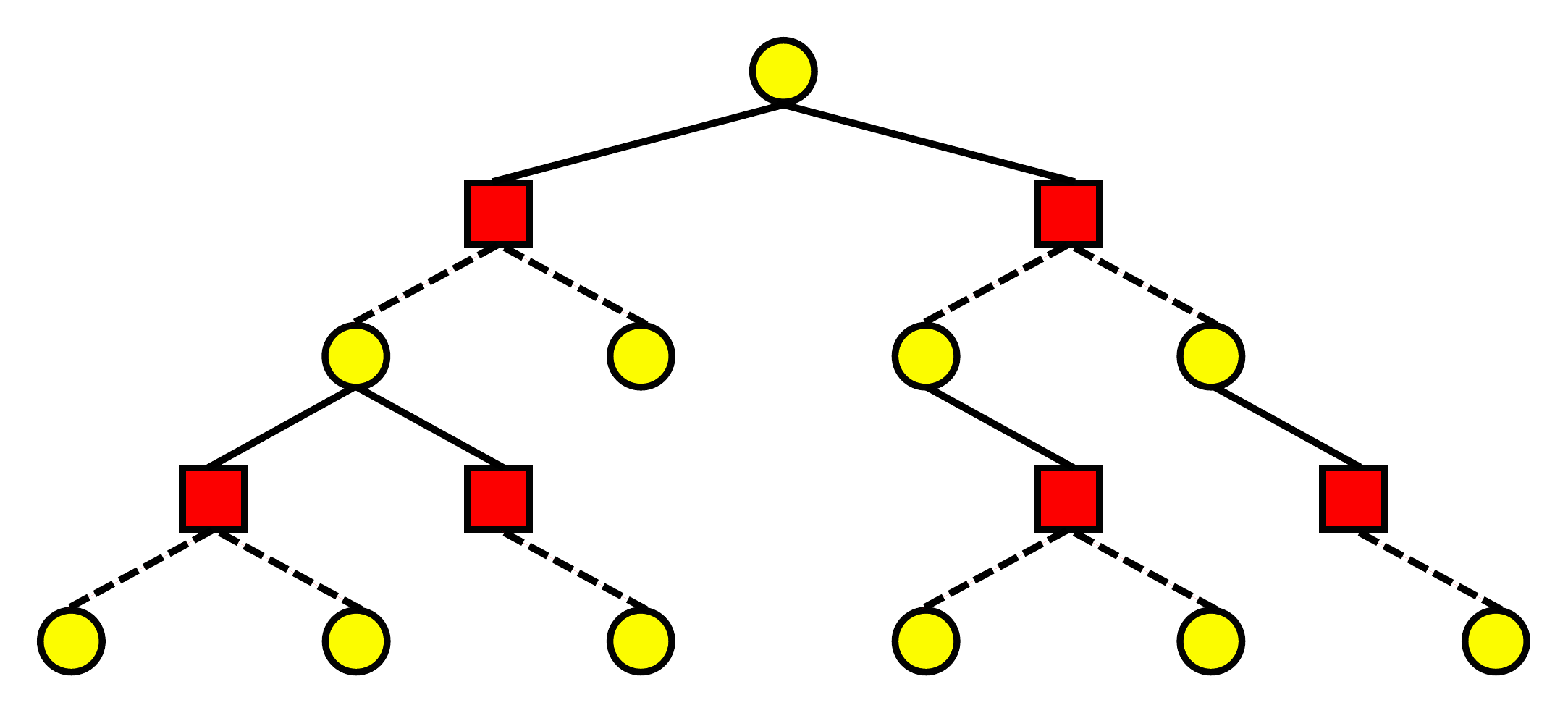}
  \caption{Small-scale example of states and actions modelled with MCTS. Yellow circles correspond to states and red squares represent the actions chosen by bandit $b_i$ of agent $i$. Dashed lines indicate non-determinism from $i$'s perspective, because the next state is determined by all agents and the environment.}
  \label{fig:vis:mcts}
  \Description{Example of MCTS with a branching factor of 2. The first node is the tree root and represents the current state. Its children are the actions of the agent which are followed by the next state. Because state transition depends on the joint action of all agents, the edges leading from action nodes to state nodes are non-deterministic.}
\end{figure}

Our study focuses on Monte Carlo Tree Search (MCTS) which selects actions through lookahead search. The search tree is built from $s_t \in \mathcal{S}$ as state nodes and $a_{i,t} \in \mathcal{A}_{i,t}$ as action nodes, starting from the current state as the root with $t = t_0$ (s. Fig. \ref{fig:vis:mcts}). Edges from $a_{i,t}$ to $s_{t+1}$ are given by $\mathcal{P}$.

Tree search is done iteratively in four steps:

\begin{itemize}
	\item \emph{Selection}: Starting from the root, $a_{i,t}$ is chosen according to the selection strategy $\pi_{\textit{tree}, i} (s_t, N)$ which is parameterized by a state $s_t$ and visit count $N$ of the state node. Transitions to the next state node $s_{t+1}$ are sampled via $\mathcal{P}$.
	\item \emph{Expansion}: When a new non-terminal state is reached with $\mathcal{P}$ and $t < h + t_0$, a corresponding state node is created and added to the last $a_{i,t}$ node's children. The new state node also creates its action node children for $a_{i,t+1} \in \mathcal{A}_{i,t}$.
	\item \emph{Rollout}: The last state node is evaluated with a rollout policy $\pi_\textit{roll}$ until $t = h + t_0$ or a terminal state is reached. In our study, $\pi_\textit{roll}$ randomly selects actions with uniform probability.
	\item \emph{Backpropagation}: $G_t$ is computed recursively from leaf to root and $\pi_{\textit{tree}, i} (s_t, N)$ is updated accordingly.
\end{itemize}

In our case, the selection strategy is set to the agent policy: $\pi_{\textit{tree}, i} (s_t, N_\textit{max}) = \pi_i (s_t)$ at all nodes where $N_\textit{max}$ is the maximum number of simulations as given by $c_b$. In this paper, we assume that a constant $c_b$ also leads to constant $N_\textit{max}$ regardless of actual runtime. Thus, for simplicity reasons, we assume $c_b=N_\textit{max}$. $\mathcal{P}$ and $\mathcal{R}$ are provided by $\mathcal{\hat{M}}$.

In our multi-agent setting, there are $n$ multi-armed bandits at each state node $s_t$ representing the joint policy $\pi$ and bandit $b_i$'s arms correspond to the actions available to agent $i$. The bandit $b_i$ represents $\pi_{\textit{tree}, i} (s_t)$ and optimizes the expected discounted return, based on $G_{i,t}$ and node visit counts $N$. Note that each agent $i$ runs MCTS locally, meaning that $b_i$ selects the own actions and $b_j$ models other agents or opponents for $j \in \{1, ..., n\} \setminus i$. All $b_j$ are used to predict the complete joint action $a_t$ which is needed for $\mathcal{P}$ and $\mathcal{R}$. Bandits are not shared between agents and thus $b_j$ is used by agent $i$ to estimate $\pi_j$ and $V_j$.

UCB1 is often used to implement $\pi_{\textit{tree}, i}$ in state-of-the-art approaches \cite{alphago, alphagozero, muzero, kocsis, bubeck}, leading to the popular \emph{Upper Confidence bound applied to Trees (UCT)} algorithm \cite{kocsis, alphago, alphagozero, muzero}.

\subsection{Searching with Opponent-Awareness}

\begin{algorithm}
\caption{OGA bandit algorithm}
\begin{algorithmic}[1]

\Procedure{sample}{$b$}
	\State Initialize action distribution $P_N$
    \For{$a_\textit{MAB} \in \mathcal{A}_\textit{MAB}$}
        \State Calculate $P_N(a)$ as in Eq. \ref{eq:grabprob}
    \EndFor
    \State $a_\textit{sample} \sim P_N$
    \State\Return{$a_\textit{sample}$}
\EndProcedure

\Procedure{update\_expectations}{$b,G_{i,t}$}
	\State $X_i \leftarrow (X \cdot N + G_{i,t}) / (N + 1)$
	\State $\bar{X}_{i,a} \leftarrow (\bar{X}_{i,a} \cdot N_a + G_{i,t}) / (N_a + 1)$
	\State $N \leftarrow N + 1$
	\State $N_a \leftarrow N_a + 1$
	\State\Return{$X_i,\bar{X}_{i,a}$}
\EndProcedure

\Procedure{gradients}{$b,X'$,$\bar{X'}_a$}
    \State $\nabla_{\theta^i_t} V(\theta_t)) \leftarrow \alpha (\bar{X'}_a - X') (\mathds{1}_{\ddot{a}_{MAB}=a_{MAB}} - P_N(\ddot{a}_{MAB}))$
    \State\Return{$\nabla_{\theta^i_t} V(\theta_t))$}
\EndProcedure

\Procedure{update\_preferences}{$b,\nabla_{\theta_t} V(\theta_t))$}
	\State $LOLA_{add} \leftarrow \sum_{j \in \{1,..., n\} \setminus i}^{} (\nabla_{\theta^j_t} V_i(\theta_t))^T \nabla_{\theta^i_t} \nabla_{\theta^j_t} V_j(\theta_t) \delta_i \delta_j$
    \State $H_{N+1} (\ddot{a}_{MAB}) \leftarrow H_N(\ddot{a}_{MAB}) + \nabla_{\theta^i_t} V_i(\theta_t)) + LOLA_{add}$
\EndProcedure

\end{algorithmic}
\end{algorithm}

\begin{algorithm}
\caption{SOA}
\begin{algorithmic}[1]

\Procedure{SOA}{$h,s_{t_0},c_b,i$}
	\State Create root state node for $s_{t_0}$
	\State $N_{MAX} \leftarrow c_b$
	\State $N \leftarrow 0$
    \While{$N < N_{MAX}$}
    	\State $isNew \leftarrow N = 0$
        \State \emph{SIMULATE\_STATE}($s_{t_0},h,isNew,i$)
        \State $N \leftarrow N + 1$
    \EndWhile
    \State $b \leftarrow b_i$ from $s_{t_0}$
    \State\Return{$sample(b)$}
\EndProcedure

\Procedure{simulate\_state}{$s_t,h,isNew,i$}
    \If{$h \leq 0$}
        \State\Return{0}
    \EndIf
    \If{$isNew$}
    	\State Create action node children \textbf{for} $a_{i,t} \in \mathcal{A}_{i,t}$
    	\State Create bandits $b_j$ \textbf{for} $j \in \{1, ... ,n\}$
    	\State Perform rollout with $\pi_\textit{roll}$ to sample $G_t$
        \State\Return{$G_t$}
    \EndIf
    \State $a_t \leftarrow sample(b_j)$ \textbf{for} $j \in \{1, ..., n\}$
	\State $G_t \leftarrow SIMULATE\_ACTION (a_t,s_t,h,i)$
	\State $UPDATE\_BANDITS(G_t)$
	\State\Return{$G_t$}
\EndProcedure

\Procedure{simulate\_action}{$a_t,s_t,h,i$}
	\State $s_{t+1},r_t \sim \mathcal{\hat{M}} (s_t,a_t)$
    \If{no state node for $s_{t+1}$ in $a_{i,t}$'s children}
    	\State Make new state node child for $s_{t+1}$
    	\State $R_t \leftarrow SIMULATE\_STATE(s_{t+1},h-1,True,i)$
        \State\Return{$r_t + \gamma R_t$}
    \EndIf
    \State $R_t \leftarrow SIMULATE\_STATE(s_{t+1},h-1,False,i)$
	\State\Return{$r_t + \gamma R_t$}
\EndProcedure

\Procedure{update\_bandits}{$G_t$}
	\State $X',\bar{X'}_a \leftarrow UPDATE\_EXPECTATIONS(b_j,G_{j,t})$ \textbf{for} $j \in \{1, ..., n\}$
	\State $\nabla_{\theta_t} V(\theta_t) \leftarrow GRADIENTS(b_j,X,\bar{X}_a)$ \textbf{for} $j \in \{1, ..., n\}$
	\State $UPDATE\_PREFERENCES(b_j,\nabla_{\theta_t} V(\theta_t))$ \textbf{for} $j \in \{1, ..., n\}$
\EndProcedure

\end{algorithmic}
\end{algorithm}

In \emph{Searching with Opponent-Awareness (SOA)}, we leverage the concept of LOLA to improve social interactions between planning agents. LOLA itself is designed for learning agents, as opposed to a planning agent. However, bandits in the search tree actually represent \emph{learning} instances and thus we develop an opponent-aware bandit which can be integrated into planning algorithms like MCTS.

The key idea behind LOLA is to incorporate the learning step of another agent into one's own update, in contrast to a naive learner which ignores the updates of other learners in their own update \cite{lola}. For a naive learning agent $i$ whose policy $\pi_i$ is parameterized by $\theta^i_t$, the update rule is defined by:

\begin{eqnarray}\label{eq:naiveupdate}
\theta^i_{t+1} = \theta^i_t + \nabla_{\theta^i_t} V_i(\theta_t) \delta_i
\end{eqnarray}

$V_i(\theta_t)$ is the state value function approximation of agent $i$ as a function of all agents' policy parameters $\theta_t = (\theta^1_t, ..., \theta^n_t)$, $\delta_i$ is the learning rate of agent $i$. LOLA adds the following term to the update \cite{lola}:

\begin{eqnarray}\label{eq:lolaupdateaddition}
LOLA_{add} = \sum_{j \in \{1,..., n\} \setminus i}^{} (\nabla_{\theta^j_t} V_i(\theta_t))^T \nabla_{\theta^i_t} \nabla_{\theta^j_t} V_j(\theta_t) \delta_i \delta_j
\end{eqnarray}

As such, the total LOLA update is given by \cite{lola}:

\begin{eqnarray}\label{eq:lolaupdate}
\theta^i_{t+1} = \theta^i_t + \nabla_{\theta^i_t} V_i(\theta_t) \delta_i + LOLA_{add}
\end{eqnarray}

Because the UCB1 update is not gradient-based, it cannot be adjusted as in Eq. \ref{eq:lolaupdate} which is intended for learning agents using policy gradient updates. However, the GRAB bandit maximizes the expected return through stochastic gradient ascent and can thus directly implement the LOLA update rule. We refer to this variant as Opponent-Gradient Aware (OGA) bandit that can be integrated into MCTS as $\pi_{\textit{tree}}$ and to simulate $\pi$ at each state node. In this way, we realize opponent-aware planning without an explicit or a priori opponent model. The complete formulation for the OGA bandit algorithm is given in Algorithm 1, where $X' = \{X_1, ..., X_n\}$, $\bar{X'}_a = \{\bar{X}_{1,a}, ..., \bar{X}_{n,a}\}$ and $\nabla_{\theta_t} V(\theta_t)$ represents all $\nabla_{\theta^i_t} V_j(\theta_t)$ with $i, j \in \{1, ..., n\}$.

SOA uses OGA bandits in the \emph{Selection} step of MCTS to sample $a_{j,t}$ as $a_\textit{MAB}$ for all $b_j$ where $j \in \{1, ..., n\}$ at each $s_t$ to obtain $s_{t+1}$ and $r_t$ from $\mathcal{\hat{M}}$. During \emph{Backpropagation}, the joint discounted return $G_t$ is used to recursively update all bandits in the path from leaf to root. The complete formulation for SOA is given in Algorithm 2. Note $X', \bar{X'}_a$ and $\nabla_{\theta_t} V(\theta_t)$ are accumulated from $X_i,\bar{X}_{i,a}$ and $\nabla_{\theta^i_t} V(\theta_t)$ for \textit{all} $i \in \{1, ..., n\}$. The \emph{UPDATE\_BANDITS} procedure can be easily adapted for other bandits like UCB1 and GRAB while leaving out opponent-awareness. We refer to the latter case as \emph{GRAB-MCTS}.

\section{Experiments}

We compare SOA, UCT and GRAB-MCTS in different settings and evaluate their performance. The first setting includes iterated matrix games: chicken drive, prisoners' dilemma and matching pennies. In these games, each agent chooses one of two actions and immediately receives a reward, based on the joint action. Afterwards, the agents are studied in coin game and predator-prey which are more complex than iterated matrix games and require a sequence of actions until a reward is obtained.

In all of our experiments, we set $\gamma = 0.9$ and $C = 1.0$. In all of our settings, only one type of bandit is used by the agents in a given episode. Episodes are sequences of time steps of length $T$ where state transitions are determined by the environment and the actions of all agents. In all of our domains, we set $T = 50$.

\subsection{Iterated matrix games}
The iterated matrix games we study in this paper represent basic challenges in multi-agent systems as agent behavior directly affects the returns of both agents. The payoff matrices for each game are shown in Tables \ref{tab:payipd}, \ref{tab:payimp} and \ref{tab:payicd}.

\begin{table}[h]
  \caption{Payoff matrix of Prisoners' Dilemma}
  \label{tab:payipd}
  \begin{tabular}{|c|c|c|}
    \hline
     & C & D \\
    \hline
    C & (-1, -1) & (-3, 0) \\
    \hline
    D & (0, -3) & (-2, -2) \\
    \hline
  \end{tabular}
\end{table}

The Iterated Prisoners' Dilemma (IPD) gives an agent the choice between cooperating (C) and defecting (D). Under the assumption of infinite iterations, there are many Nash Equilibria as shown by the folk theorem \cite{myersongametheory}, including mutual defection and Tit-for-Tat (TFT) with expected returns of -2 and -1 respectively. The latter strategy begins by first cooperating and then imitating the opponent's previous move.

\begin{table}[h]
  \caption{Payoff matrix of Matching Pennies}
  \label{tab:payimp}
  \begin{tabular}{|c|c|c|}
    \hline
     & H & T \\
    \hline
    H & (+1, -1) & (-1, +1) \\
    \hline
    T & (-1, +1) & (+1, -1) \\
    \hline
  \end{tabular}
\end{table}

Iterated Matching Pennies (IMP) is a zero-sum game whose only mixed strategy Nash Equilibrium is for both agents to play head (H) or tail (T) with a probability of 50\%, leading to expected returns of 0.

\begin{table}[h]
  \caption{Payoff matrix of Chicken Drive}
  \label{tab:payicd}
  \begin{tabular}{|c|c|c|}
    \hline
     & C & D \\
    \hline
    C & (0, 0) & (-1, +1) \\
    \hline
    D & (+1, -1) & (-10, -10) \\
    \hline
  \end{tabular}
\end{table}

In Iterated Chicken Drive (ICD), agents have the option to chicken (C) or to drive (D). This game also has many different Nash Equilibria, most notably two pure strategies where one agent always chooses C and the other selects D with expected returns of -1 and +1 and vice versa.

\subsection{Coin game}

\begin{figure}
  \centering
  \includegraphics[width=\linewidth]{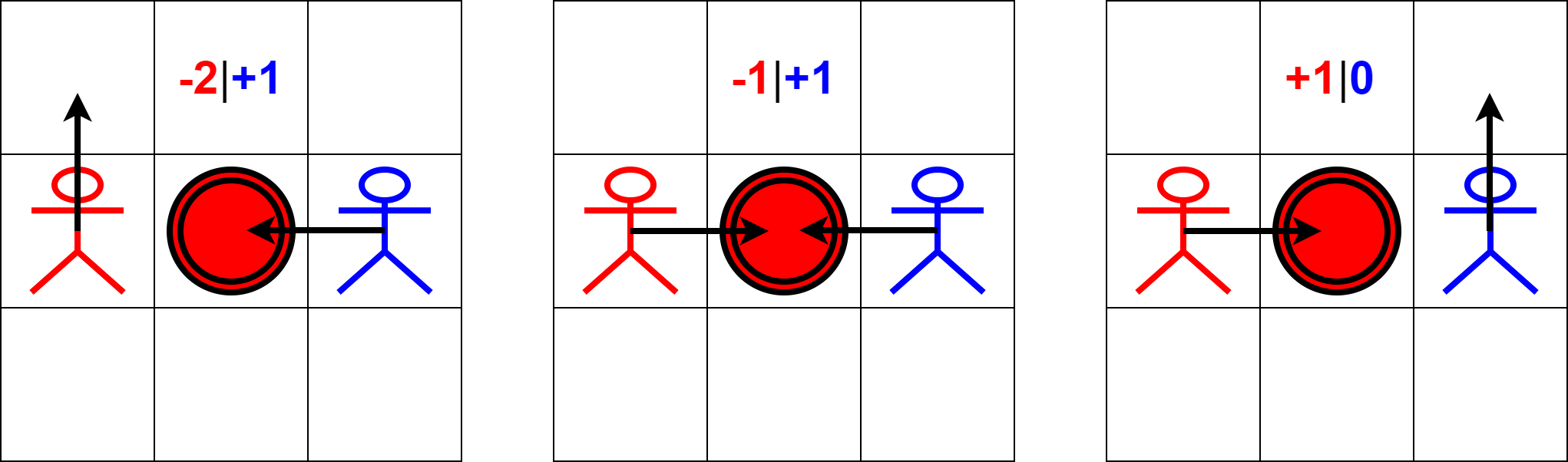}
  \caption{Example states for coin game with a red coin. Picking up any coin yields a reward of +1, but the red agent is penalized with -2 if a red coin is picked up by the blue agent and vice versa.}
  \label{fig:vis:coingame}
  \Description{Three possible outcomes of a state where the agents have the opportunity to pick up a red coin. The rewards for each agent are listed when only one or both agents pick up the coin.}
\end{figure}

Coin game is a sequential game and more complex than iterated matrix games, proposed by Lerer and Peysakhovich as an alternative to IPD \cite{lerer}. In coin game, the objective for both agents is to collect coins in a grid-world by moving to the coin's position. There is one red and one blue agent and coins are either red or blue.

Picking up any coin yields a reward of +1 point, but if e.g. the blue agent picks up a red coin, the red agent is penalized with a reward of -2 (and vice versa). A simple greedy policy yields an expected return of 0 \cite{lerer, lola}.

\subsection{Predator-prey}

\begin{figure}
  \centering
  \includegraphics[width=\linewidth]{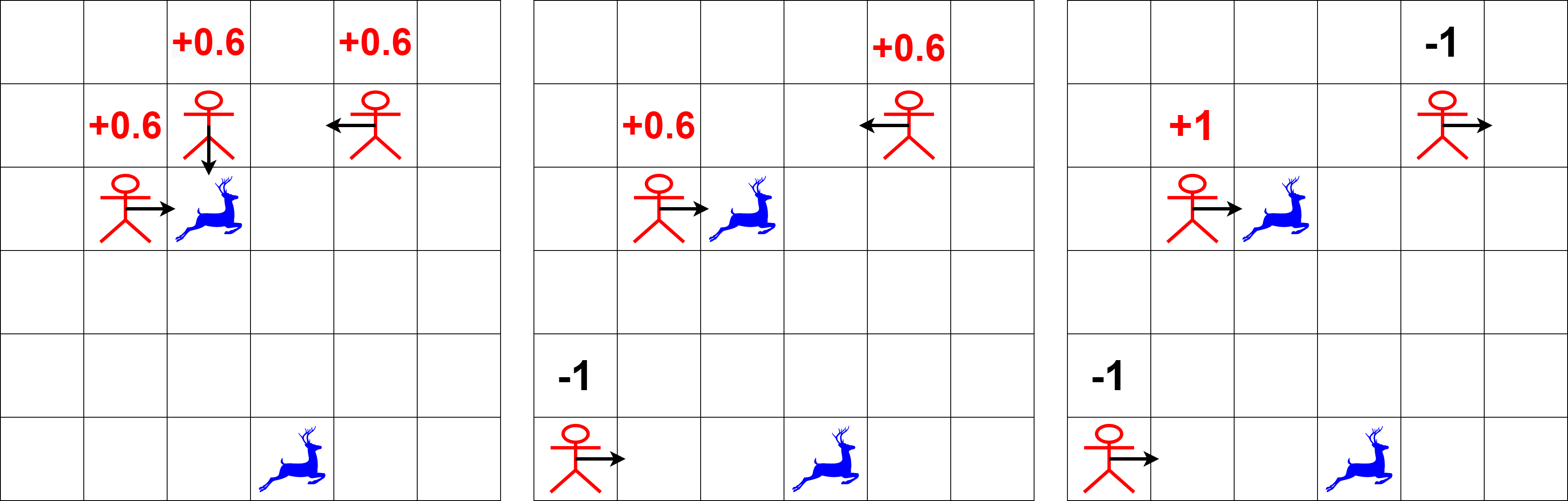}
  \caption{Example states for predator-prey with 3 agents (red) and 2 uncaptured preys (blue). Capturing a prey yields a reward of 1 if only one agent is involved, 0.6 for each agent otherwise. Excluded predators suffer hunger, represented by a -1 penalty.}
  \label{fig:vis:pre}
  \Description{Three possible state transitions where a prey is captured. The rewards and penalties are shown when all agents share the captured prey, when an agent is excluded and when only one agent captures the prey.}
\end{figure}

In predator-prey, agents seek to catch preys in a grid-world. Unlike in coin game, preys are mobile targets and not respawned when caught. An episode ends when all preys have been captured or $T = 50$ time steps have elapsed.

Agents receive a reward of +1 when capturing a prey alone. When at least one more agent has a Chebyshev distance of 1 or less, the prey is shared among these agents with a reward of +0.6 for each agent. Agents who are farther away are penalized with a reward of -1 to simulate hunger.

\section{Results}

\subsection{Iterated matrix games}

\begin{figure*}
	\centering
	\begin{subfigure}[b]{0.3\textwidth}
		\centering
		\includegraphics[scale=0.36]{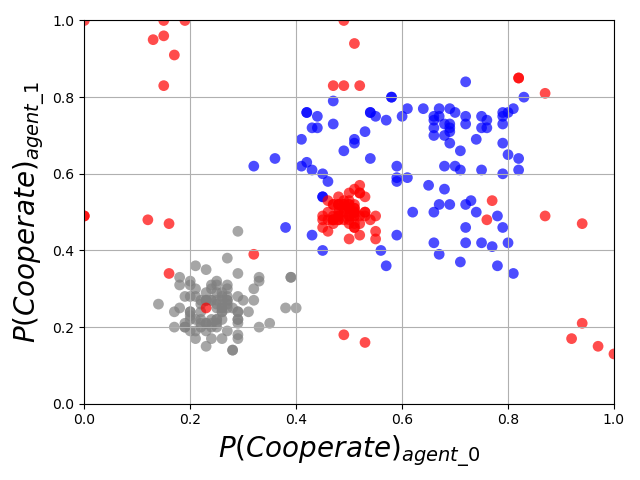}
		\caption{}
		\label{fig:res:img:ipd}
	\end{subfigure}
	\quad
	\begin{subfigure}[b]{0.3\textwidth}
		\centering
		\includegraphics[scale=0.36]{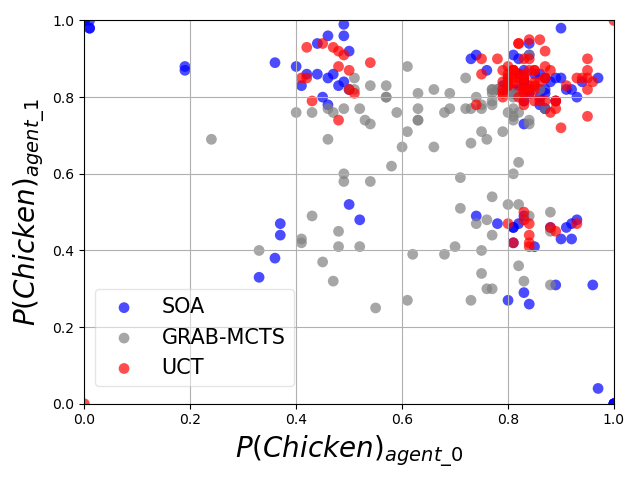}
		\caption{}
		\label{fig:res:img:icd}
	\end{subfigure}
	\quad
	\begin{subfigure}[b]{0.3\textwidth}
		\centering
		\includegraphics[scale=0.36]{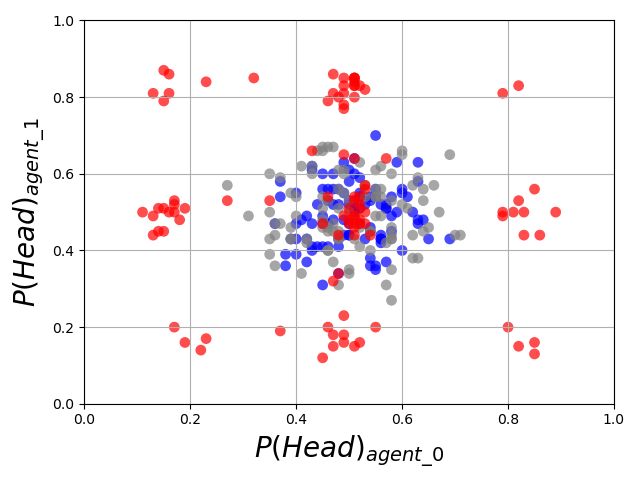}
		\caption{}
		\label{fig:res:img:imp}
	\end{subfigure}
	\caption{Distribution of relative action count per episode for both agents across 100 episodes. (a) Probability of cooperating in IPD. (b) Probability of avoiding collision in ICD. (c) Probability of selecting head in IMP.}
	\Description{Scatterplots where a point's coordinates are the probability of each agent selecting one action throughout the entire episode. The colors of the scattered points correspond to one selection strategy each. In IPD, GRAB-MCTS leads to mutual defection, UCT to near-random behavior and SOA to cooperation. In ICD, all algorithms converge mostly to Chicken, though UCT and SOA sometimes alternate between Chicken and Drive when the opponent is likely to Chicken. In IMP, all algorithms converge mainly to the Nash equilibrium.}
	\label{fig:res:img}
\end{figure*}

In iterated matrix games, the policy of an agent can be conditioned on past $d$ states, though Press and Dyson have proven that remembering one past state is sufficient \cite{pressdyson}. Thus, we set the horizon $h = 2$ and model the initial state with the root. The policy for subsequent states is derived from the previous state transition which is taken as the tree path from the root to the bandit who selects the next action. $N_{max}$ is kept constant at 100.

Fig. \ref{fig:res:img} illustrates the distribution of chosen actions for each selection strategy. In IPD, the use of SOA results in a higher probability of mutual cooperation compared to UCT and GRAB-MCTS, shown by the relative distances from the three pointclouds to the upper right corner in Fig \ref{fig:res:img:ipd}.

We analyze the agents in the same way in ICD and find that all agents converge to mostly playing Chicken, as shown by the concentration of points in the upper right quadrant of Fig. \ref{fig:res:img:icd}. UCT has the most stable selection strategy in this setting and GRAB-MCTS leads to the highest variance in the probability of playing Chicken. The small groups near the upper middle and the middle right indicate that UCT and SOA agents also have a tendency of playing Drive when the opponent is likely to play Chicken.

In IMP, SOA and GRAB-MCTS converge to the mixed strategy Nash equilibrium where at least one agent's policy approximates uniform distribution of actions, evidenced by the cluster in the center of Fig \ref{fig:res:img:imp}. UCT sometimes develops a preference for playing either heads or tails, shown by the concentrations which surround the center. Nevertheless, if the opponent uses the mixed strategy Nash equilibrium, the expected return is still 0.

\subsection{Coin game}

\begin{figure*}
	\centering
	\begin{subfigure}[b]{0.3\textwidth}
		\centering
		\includegraphics[scale=0.35]{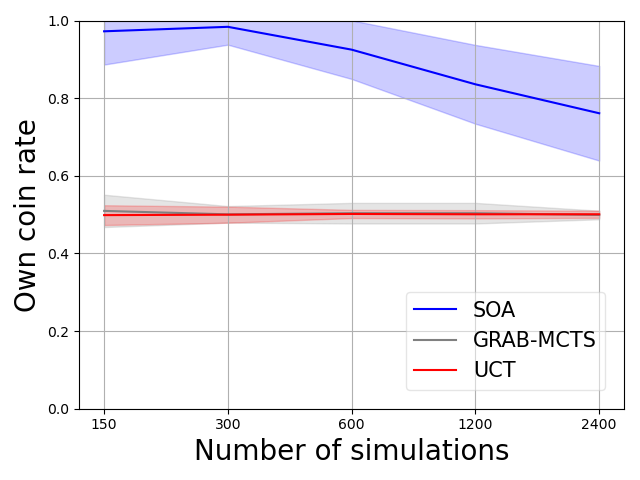}
		\caption{}
		\label{fig:res:coin:own}
	\end{subfigure}
	\quad
	\begin{subfigure}[b]{0.3\textwidth}
		\centering
		\includegraphics[scale=0.35]{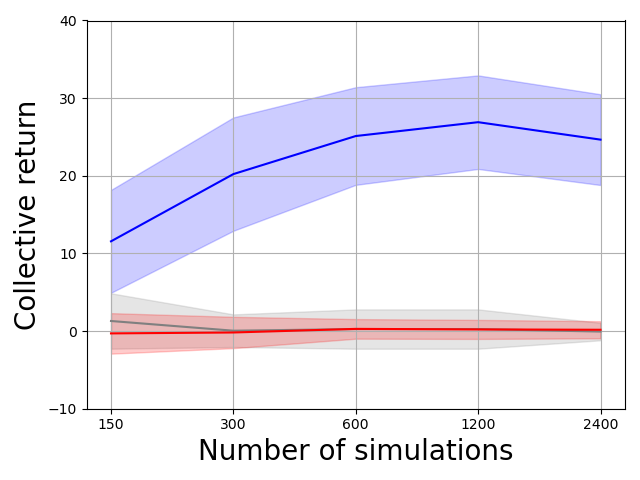}
		\caption{}
		\label{fig:res:coin:ret}
	\end{subfigure}
	\quad
	\begin{subfigure}[b]{0.3\textwidth}
		\centering
		\includegraphics[scale=0.35]{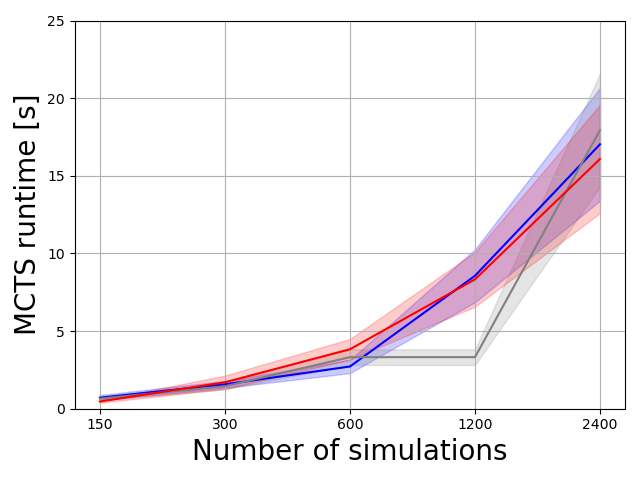}
		\caption{}
		\label{fig:res:coin:run}
	\end{subfigure}
	\caption{Performance comparison of all three planning algorithms in coin game across 50 episodes per algorithm type per budget. Shaded areas show standard deviation. (a) Probability for each agent to pick up a coin of their own color. (b) Collective undiscounted return $W$ per algorithm type. (c) MCTS runtime per agent per time step.}
	\label{fig:res:coin}
	\Description{Line graphs showing the performance of the planning algorithms in coin game. The y-axis represent the descriptions in the caption and the x-axis is the number of simulations. SOA agents are consistently more likely to pick up coins of their own color and have higher returns than the other planning agent whereas UCT and GRAB-MCTS agents behave similarly. All algorithms show similar runtime behavior.}
\end{figure*}

In coin game, we examine the behavior of agents in a sequential game where the exact impact of single actions of a cooperative or defective strategy on the future return is less obvious than a reward matrix. In this setting, we also compare the computational scalability of all three selection strategies by measuring the time needed to consume their computation budget. The gridworld size is $3 \times 3$ and horizon $h = 6$.

Fig. \ref{fig:res:coin:own} shows the probability of an agent picking up coins of the same color as their own. Both GRAB-MCTS and UCT lead to a mean probability of about 50 \%, meaning that these agents pick up coins regardless of its color. SOA agents are more likely to pick up coins of their own color, though this probability decreases with increased $N_{max}$. However, as shown in Fig \ref{fig:res:coin:ret}, this decline is not reflected in the collective return $W$ (s. Eq. \ref{eq:return:undiscounted:collective}) which actually increases. This is due to the facts that a coin is spawned if and only if a coin has ben picked up and that we set a time limit $T = 50$ in our episodes. Therefore, the maximum number of coins to be picked up is 50 (if a coin is picked up at every time step) and decreases with every time step where one agent waits for their opponent to reach their coin. This presents a trade-off between looking for one's own coin and forcing coin respawns.

As shown in Fig. \ref{fig:res:coin:run}, MCTS runtimes with all three bandits are comparable in this setting, even when $c_b$ varies.

\subsection{Predator-prey}

\begin{figure*}
	\centering
	\begin{subfigure}[b]{0.3\textwidth}
		\centering
		\includegraphics[scale=0.35]{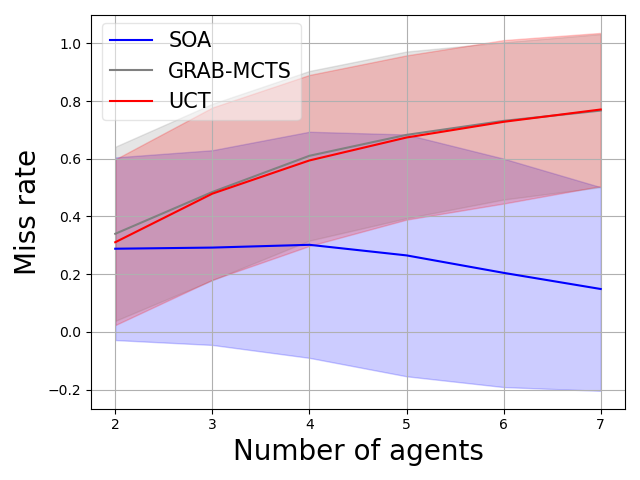}
		\caption{}
		\label{fig:res:pre:0:miss}
	\end{subfigure}
	\quad
	\begin{subfigure}[b]{0.3\textwidth}
		\centering
		\includegraphics[scale=0.35]{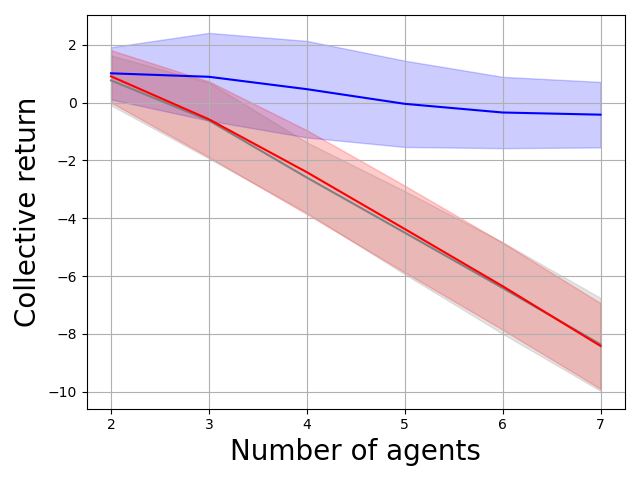}
		\caption{}
		\label{fig:res:pre:0:ret}
	\end{subfigure}
	\quad
	\begin{subfigure}[b]{0.3\textwidth}
		\centering
		\includegraphics[scale=0.35]{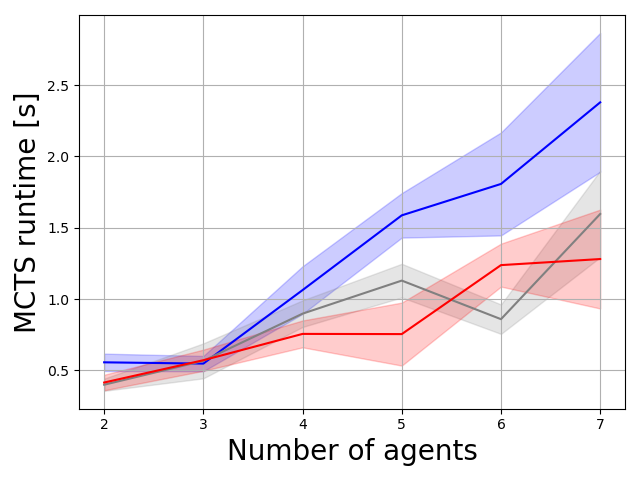}
		\caption{}
		\label{fig:res:pre:0:run}
	\end{subfigure}
	\quad
	\begin{subfigure}[b]{0.3\textwidth}
		\centering
		\includegraphics[scale=0.35]{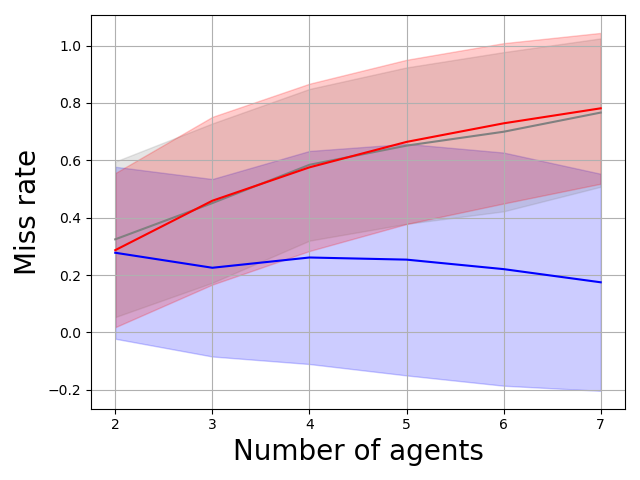}
		\caption{}
		\label{fig:res:pre:1:miss}
	\end{subfigure}
	\quad
	\begin{subfigure}[b]{0.3\textwidth}
		\centering
		\includegraphics[scale=0.35]{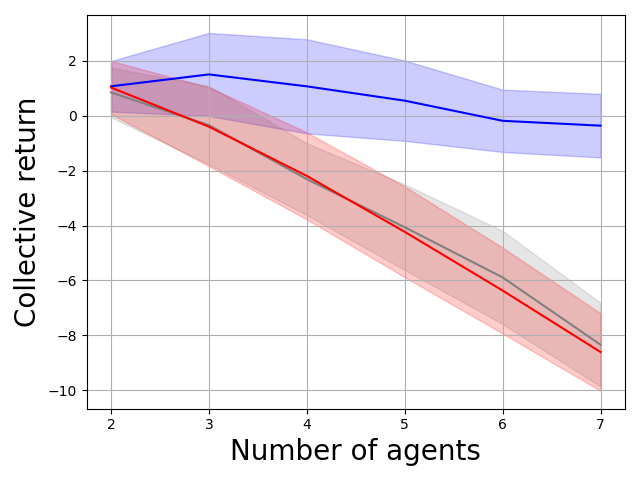}
		\caption{}
		\label{fig:res:pre:1:ret}
	\end{subfigure}
	\quad
	\begin{subfigure}[b]{0.3\textwidth}
		\centering
		\includegraphics[scale=0.35]{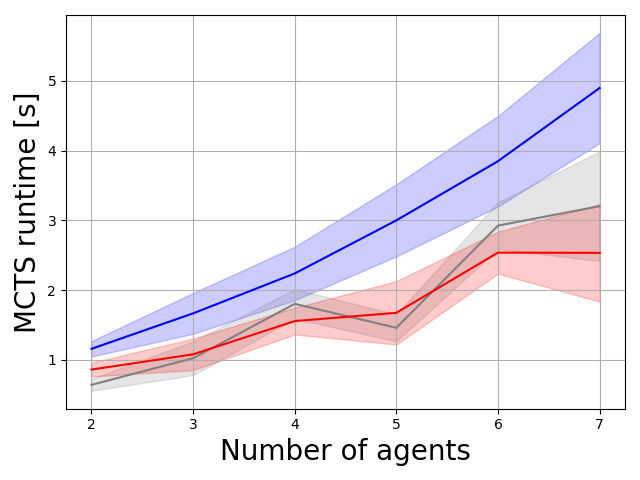}
		\caption{}
		\label{fig:res:pre:1:run}
	\end{subfigure}
	\quad
	\begin{subfigure}[b]{0.3\textwidth}
		\centering
		\includegraphics[scale=0.35]{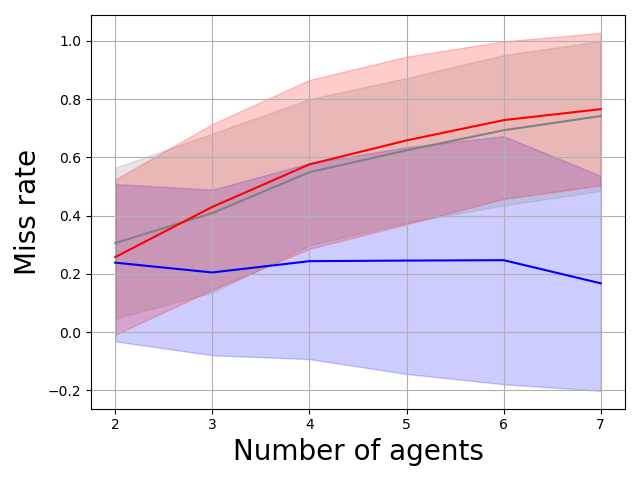}
		\caption{}
		\label{fig:res:pre:2:miss}
	\end{subfigure}
	\quad
	\begin{subfigure}[b]{0.3\textwidth}
		\centering
		\includegraphics[scale=0.35]{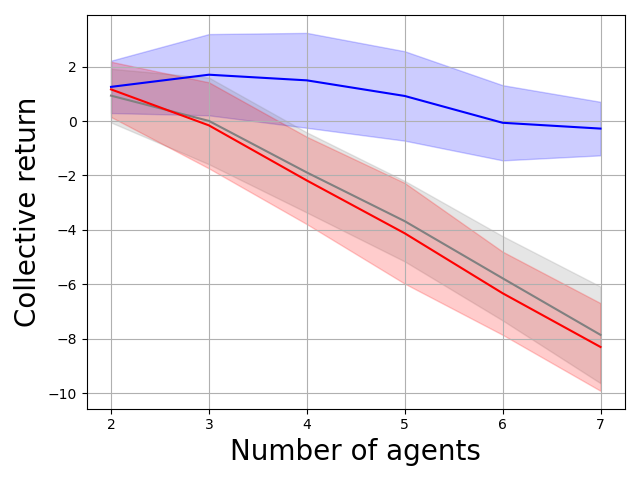}
		\caption{}
		\label{fig:res:pre:2:ret}
	\end{subfigure}
	\quad
	\begin{subfigure}[b]{0.3\textwidth}
		\centering
		\includegraphics[scale=0.35]{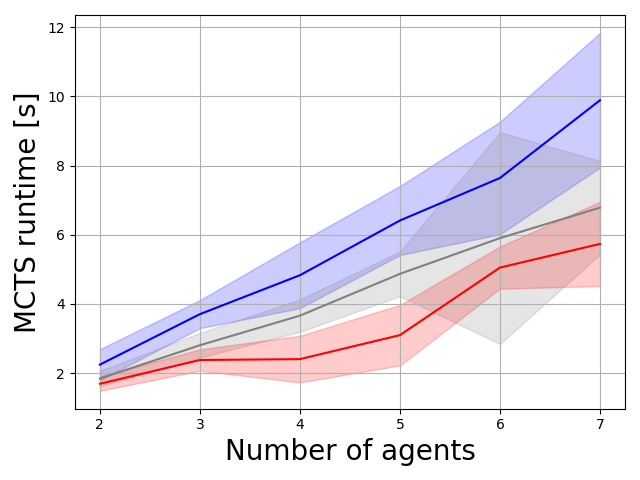}
		\caption{}
		\label{fig:res:pre:2:run}
	\end{subfigure}
	\caption{Performance comparison of all three planning algorithms in predator-prey across 100 episodes per algorithm per number of agents. Shaded areas show standard deviation. (a, b, c) compare performance with $l = 1$, (d, e, f) with $l = 2$ and (g, h, i) with $l = 4$ ($N_\textit{max} = 50 \cdot l \cdot (n + 3)$). (a, d, g) Probability for each agent to be excluded from the capture of a prey. (b, e, h) Collective discounted return $W$ per MCTS type. (c, f, i) MCTS runtime per agent per time step.}
	\label{fig:res:pre}
	\Description{Line graphs showing the performance of the planning algorithms in predator-prey. The y-axis represent the descriptions in the caption and the x-axis is the number of agents. SOA agents are consistently less likely to be excluded from the capture of prey, have higher returns than the other planning agents and SOA has the highest runtime of all planning algorithms. UCT and GRAB-MCTS show similar behavior.}
\end{figure*}

In predator-prey, we also test the scalability of the bandit (and by extension planning) algorithms w.r.t. the number of agents with $n \in \{2, 3, 4, 5, 6, 7\}$ while keeping the number of prey constant at 2. We further set the horizon $h = 2 g$, the gridworld size to $g \times g$ and $N_\textit{max} = 50 \cdot l \cdot g$ where $g = n + 3$ and $l \in \{1, 2, 4\}$. For example, if $n=3$ and $l=1$, $h=12, N_\textit{max} = 300$ and the gridworld has size $6 \times 6$. If $n=5$ and $l=2$, $h=16, N_\textit{max} = 800$ and the gridworld has size $8 \times 8$.

Fig. \ref{fig:res:pre} displays the accumulated results of agent behavior, collective returns $W$ (see Eq. \ref{eq:return:undiscounted:collective}) for each episode and MCTS runtime per agent per time step. UCT and GRAB-MCTS agents simply hunt down prey, leading to a high probability for an agent to be penalized with increasing $n$ due to preys being limited to 2 instances. This is also reflected in declining collective returns $W$ because more agents are penalized.

In contrast, SOA leads to consistently low probability of an agent being penalized and higher collective returns $W$. This is due to an agent's tendency to avoid prey until other agents are in close proximity. Increasing $n$ makes this more difficult, because it requires coordination between more agents. Additionally, preys are moving targets which means that staying still means to risk that preys escape and that increasing $n$ (which also increases the gridworld size) leads to a larger state space. Note that not catching any prey results in a return of 0 for all agents, which yields a higher $W$ than when many agents are penalized due to defective captures.

This observation can also be explained by examining the difference between naive learning in Eq. \ref{eq:naiveupdate} and the LOLA update rule in Eq. \ref{eq:lolaupdate} which is the $LOLA_{add}$ term defined in Eq. \ref{eq:lolaupdateaddition}. In Eq. \ref{eq:lim:naive}, we observe the impact of increasing $n$ on a naive update and find that the term remains constant.

\begin{equation} \label{eq:lim:naive}
\lim \limits_{n \to \infty} \theta^i_t + \nabla_{\theta^i_t} V_i(\theta_t) \delta_i = \theta^i_t + \nabla_{\theta^i_t} V_i(\theta_t) \delta_i
\end{equation}

However, increasing $n$ causes $LOLA_{add}$ to grow arbitrarily large, as shown in Eq. \ref{eq:lim:lola}.

\begin{align} \label{eq:lim:lola}
& \lim \limits_{n \to \infty} & LOLA_{add} & = & \notag\\
& \lim \limits_{n \to \infty} & \sum_{j \in \{1,..., n\} \setminus i}^{} (\nabla_{\theta^j_t} V_i(\theta_t))^T \nabla_{\theta^i_t} \nabla_{\theta^j_t} V_j(\theta_t) \delta_i \delta_j & = \infty &\notag\\
\end{align}

Therefore, with each additional agent in a game, the naive learning component of a LOLA-learner (and by extension our OGA bandit's) is outweighed by the naive updates of all other agents. Eq. \ref{eq:lim:ratio} shows the ratio between the naive update and $LOLA_{add}$ converges to 0 with infinite $n$.

\begin{align} \label{eq:lim:ratio}
& \lim \limits_{n \to \infty} & \frac{\theta^i_t + \nabla_{\theta^i_t} V_i(\theta_t) \delta_i}{LOLA_{add}} & = & \notag\\
& \lim \limits_{n \to \infty} & \frac{\theta^i_t + \nabla_{\theta^i_t} V_i(\theta_t) \delta_i}{\sum_{j \in \{1,..., n\} \setminus i}^{} (\nabla_{\theta^j_t} V_i(\theta_t))^T \nabla_{\theta^i_t} \nabla_{\theta^j_t} V_j(\theta_t) \delta_i \delta_j} & \stackrel{\ref{eq:lim:naive}, \ref{eq:lim:lola}}{=} 0 &\notag\\
\end{align}

This can be adjusted by assigning weights or by normalizing $LOLA_{add}$, though we leave this for future work.

Fig. \ref{fig:res:pre:0:run}, \ref{fig:res:pre:1:run} and \ref{fig:res:pre:2:run} show that SOA has steeper runtime demands than UCT and GRAB-MCTS with increasing $n$. This is a side-effect of its update rule which leads to additional computation per agent.

Note the results in Fig. \ref{fig:res:pre} remain stable with different budgets $N_\textit{max}$ as shown by setting different values for $l$.

\section{Conclusion}

In this paper, we presented SOA, an MCTS variant using opponent-aware bandits in MCTS to improve social interaction and performance in multi-agent planning.

For that, we introduced OGA bandits which extend gradient-based multi-armed bandits with LOLA. This is done by adapting the GRAB bandit algorithm which is not possible for UCB1 because UCB1 updates are not gradient-based.

These OGA bandits are used in SOA to implement the tree selection policy of the MCTS and to simulate the joint policy $\pi$ at each state. During \emph{Backpropagation}, the joint discounted return $G_t$ is used for opponent-aware updates of all bandits in each visited node.

To evaluate SOA, we compared its performance in several environments to other MCTS variants with different bandit algorithms. Our experiments show that planning with opponent-awareness leads to more cooperation at the cost of more expensive computation for each agent. The benefits of SOA are especially noticeable with increasing number of agents which is useful for scaling multi-agent systems.

Opponent-aware planning may help to improve social interaction between agents in real-world applications where naive decision-making can lead to undesirable consequences, like accidents involving autonomous cars or wasted resources in an industrial factory.

For the future, we intend to study SOA in hybrid approaches with deep reinforcement learning.

\bibliographystyle{ACM-Reference-Format}
\bibliography{sources}

\end{document}